%% file: paper.tex
\documentclass[useAMS,usenatbib]{mn2e}
\usepackage{graphicx}
\usepackage{amsmath}

\input{aas_macros}

\title{A comprehensive comparison of cosmological models from latest observational data}
\author[K. Shi, Y. F. Huang, and T. Lu]{K. Shi$^{1,2}$, Y. F. Huang$^{1,2}$\thanks{E-mail:hyf@nju.edu.cn} and T. Lu$^{3,4}$ \\
$^{1}$Department of Astronomy, Nanjing University, Nanjing 210093, China\\
$^{2}$Key Laboratory of Modern Astronomy and Astrophysics (Nanjing University), Ministry of Education, Nanjing 210093, China\\
$^{3}$Purple Mountain Observatory, Chinese Academy of Sciences, Nanjing 210008, China\\
$^{4}$Joint Center for Particle, Nuclear Physics and Cosmology, Nanjing University -- Purple Mountain Observatory, Nanjing  210093, China}
\begin{document}
\date{Accepted .  Received ; in original form }
\pagerange{\pageref{firstpage}--\pageref{lastpage}} \pubyear{2011}
\maketitle
\label{firstpage}
\begin{abstract}
We investigate in detail some popular cosmological models in light of the latest observational data, including the Union2.1 supernovae compilation, the baryon acoustic oscillation measurements from the WiggleZ Dark Energy Survey, the cosmic microwave background information from the WMAP 7-year observations, along with the observational Hubble parameter data. Based on the model selection statistics such as the Akaike and the Bayesian information criterias, we compare different models to assess the worth of them. We do not assume a flat universe in the fitting. Our results show that the concordance $\Lambda$CDM model remains the best one to explain the data, while the DGP model is clearly disfavored by the data. Among these models, those whose parameters can reduce themselves to the $\Lambda$CDM model provide good fits to the data. These results indicate that for the current data, there is no obvious evidence supporting any more complex models over the simplest $\Lambda$CDM model.
 
\end{abstract}


\section{Introduction}

The present accelerating expansion of the universe is a great challenge to our fundamental physics and cosmology. This fact was first discovered by Type Ia supernova (SNIa) surveys \citep{1998AJ....116.1009R,1999ApJ...517..565P}, and later confirmed by precise measurement of the Cosmic Microwave Background (CMB) anisotropies \citep{2003ApJS..148..175S} as well as the baryon acoustic oscillations (BAO) in the Sloan Digital Sky Survey (SDSS) luminous galaxy sample \citep{2005ApJ...633..560E}. This cosmic acceleration leads us to believe that most energy in the universe exists in the form of a new ingredient called ``dark energy'' which has a negative pressure.

Various theoretical models of dark energy have been proposed, the simplest being the cosmological constant $\Lambda$ with constant dark energy density and equation of state $w_{DE} = p/\rho = -1$. This model, the popular $\Lambda$CDM model, provides an excellent fit to a wide range of observational data so far. Despite its simplicity and success, the $\Lambda$CDM model has two problems. One is the so called ``fine tuning'' problem, that is, the observed value of $\Lambda$ being extremely small comparing with particle physics expectations \citep{1989RvMP...61....1W}. The other is the coincidence problem, i.e. the present energy density of dark energy $\Omega_{\Lambda}$ and the present matter density $\Omega_m$ are of the same order of magnitude, for no physical obvious reasons. Due to these difficulties of the cosmological constant,  numerous alternative models have been proposed to explain the acceleration instead of the $\Lambda$CDM model (see \citealt{2006IJMPD..15.1753C, 2008ARA&A..46..385F} for recent reviews). Generally speaking, these models can be divided into two groups: one is to modify the matter (i.e. the right-hand side of the Einstein equation) and the other is to modify the gravity (i.e. the left-hand side of the Einstein equation). 

Although most studies show that the $\Lambda$CDM model is in good agreement with observational data, dynamical dark energy cannot be excluded yet. In order to distinguish between different dark energy models from observations, the most commonly used method is to constrain dark energy equation of state $w$. Recent studies have already given tight constraints on $w$, e.g. the Supernova Legacy Survey three year sample (SNLS3) combining with other probes has given $w$ = $-$ 1.061 $\pm$ 0.068 \citep{2011arXiv1104.1444S}. It should be noticed that although these results are \textbf{consistent} with the $\Lambda$CDM model, we cannot yet determine whether the density of dark energy is actually constant or whether it varies with time as suggested by dynamical models.

When one proposes a new cosmological model, it is of great importance to place constraints on the model parameters. Usually people use a maximum likelihood estimate to set constraints on the parameters of the model. If the expected distribution of the data is Gaussian (which is applicable for most problems in cosmology), we can use the familiar $\chi^2$ test for parameter estimation -- the smaller the $\chi^2$, the better the parameters fits the data.

On the other hand, in the face of so many different dark energy models, a natural question is raised: Which model is better, or in other words, which one is most favored by the current observational data? This is the problem of model selection. One may naively apply the $\chi^2$ test here, but that does not contain the information of the complexity (the number of parameters) of different models. That is, $\chi^2$ statistics are good at finding the best-fit parameters in a model but are insufficient for deciding whether this model itself is the best one. In order to solve this problem, some model selection statistics have been proposed in the context of cosmology \citep{2004MNRAS.351L..49L, 2007ApJ...666..716D}. The most commonly used is the information criteria (IC) including the Akaike information criterion (AIC; \citealt{1974IEEE..19.716}) and the Bayesian information criterion (BIC; \citealt{1978Ann.Statist...6..461}). These criterions tend to favor models which give a good fit with fewer parameters that embody the spirit of Occam's razor: ``entities must not be multiplied beyond necessity''.

In this paper, we investigate parameter constraints on a number of cosmological models by performing a Markov Chain Monte Carlo (MCMC) analysis from latest observational data. We then apply the model selection statistics to compare different models to assess which is preferred or disfavored by the data. We organized our paper as follows. In Section 2 we discuss the model comparison statistics used in this paper. In Section 3 we describe the observational data used in this paper and the method to use them. Section 4 gives a detailed description of different cosmological models to be tested and the constraining results from observations. The comparison of different models by using model selection statistics are given in Section 5. The last section presents our discussions and conclusions.

\section{model selection statistics}
As mentioned in the introduction, we mainly use the information criteria (IC) including AIC and BIC to test different models. A detailed description of AIC and BIC can be found in \citet{2004MNRAS.351L..49L}. The AIC is given by
\begin{equation}
 \textrm{AIC} = -2 \ln \mathcal{L}_{max} + 2k,
\end{equation}
where $\mathcal{L}_{max}$ is the maximum likelihood, $k$ is the number of parameters. Note that for Gaussian posterior distribution, $\chi^2_{min}$ = $-$ 2 $\ln\mathcal{L}_{max}$. The AIC was derived from information theoretic considerations.

The BIC is defined as
\begin{equation}
\textrm{BIC} = -2 \ln \mathcal{L}_{max} + k \ln N,
\end{equation}
where $N$ is the number of data points used in the fit. The BIC is similar to the AIC, but it includes the number of data points in its form while the AIC doesn't. Note that for any likely data set $\ln N > 2$, thus the BIC imposes a stricter penalty against extra parameters than the AIC. But the AIC remains useful as it gives an upper limit to the number of parameters which should be included. The BIC was derived as an approximation to the Bayesian evidence, but this approximation is quite crude.

The preferred model is the one that minimizes the AIC and the BIC. However, the absolute values of them is not of interest, only the relative value between different models makes sense. 

For the AIC, \citet{doi:10.1198/tech.2003.s146} featured the following ``strength of evidence'' in the form of $\Delta \textrm{AIC} = \textrm{AIC}_i - \textrm{AIC}_{min}$:

$\begin{array}{cc}
\Delta \textrm{AIC} & \textrm{Level~of~Empirical~Support~For~Model~i}\\

0 - 2 & \textrm{Substantial} \\

4 - 7 & \textrm{Considerably~Less}\\

> 10 & \textrm{Essentially~None}
\end{array}$

For the BIC, \citet{1995JSTOR.90..773} featured the following ``strength of evidence'', where $\Delta \textrm{BIC} = \textrm{BIC}_i - \textrm{BIC}_{min}$:

$\begin{array}{cc}
\Delta \textrm{BIC} & \textrm{Evidence~against~Model~i}\\

0 - 2 & \textrm{Not~Worth~More~Than~A~Bare~Mention}\\

2 - 6 & \textrm{Positive}\\

6 - 10 & \textrm{Strong}\\

>10 & \textrm{Very~Strong}
\end{array}$

Thus we can first obtain a model which minimizes the ICs, and then we can compare the rest models with it using the above judgements as a ``strength of evidence''.

It should be noticed that the information criteria alone can at most indicate that a more complex model is not necessary to explain the data, since a poor information criterion might rise from the fact that the data are too poor to constrain the extra parameters in the model, and this model might be preferred with improved data.

Furthermore, we must be aware of the limitation of using these simplified ICs, since they are based on the best-fit $\chi^2$. A more in-depth analysis of model selection should consider how much parameter space would give the data with high probability, as well as the correlations between the parameters. The Bayesian evidence is an approach that takes this into account which computes the average likelihood of a model over its prior parameter ranges. See, e.g. \citet{2004MNRAS.348..603S,2007MNRAS.377L..74L, 2007MNRAS.378...72T} for more discussions. However, the Bayesian evidence needs to compute a multidimensional integration over the likelihood and prior which may be rather complicated. In this paper, we'd like to use the ICs instead of the Bayesian evidence to compare different dark energy models. This simpler approach is sufficient for our purpose.

Besides the ICs, we also apply the reduced chi-square and the goodness of fit statistics to see how well the model fit the data. The reduced chi-square is $\chi^2$/$\nu$, where $\nu$ is the degrees of freedom usually given by $N-k$. It describes how well a model fits the observational data sets. The goodness of fit (GoF) gives the probability of obtaining a larger discrepancy between the model and the data than that observed, assuming that the model is correct. It is defined as $\textrm{GoF} = \Gamma(\nu/2, \chi^2/2)/\Gamma(\nu/2)$ where $\Gamma$ is the incomplete gamma function.
  
\section{current observational data sets}
In this section, we describe the latest data sets used in this paper and the method to analyze them.

\subsection{Type Ia Supernova (SNIa)}

Currently, SNIa is the most powerful tool to study dark energy because of \textbf{their role as standardizable candles}. For the SNIa data, we use the currently largest Union2.1 compilation \citep{2011arXiv1105.3470S} \textbf{that contains a total of} \textbf{580 SNIa, which is} an updated version of the Union2 compilation \citep{2010ApJ...716..712A}. The newly added twenty supernovae are all at relative high redshift ($0.6<z<1.4$) and thus can help tighten the constraints on the evolution behavior of dark energy.   

Cosmological constraints from SNIa data are obtained through the distance modulus $\mu(z)$. The theoretical distance modulus is
\begin{equation}
\mu_{th}(z_i) = 5 \log_{10} D_L(z_i) + \mu_0,
\end{equation}
 where $\mu_0 = 42.38-5\log_{10}h$ with $h$ the Hubble constant $H_0$ in units of 100 km/s/Mpc, and the Hubble-free luminosity distance $D_L$ is defined as
\begin{equation}
D_L(z) = \frac{1+z}{\sqrt{\vert\Omega_k\vert}}\textrm{sinn}\left[\sqrt{\vert\Omega_k\vert}\int_0^z \frac{dz'}{E(z')}\right],
\end{equation}
where $E(z) = H(z)/H_0$ and $\Omega_k$ is the present curvature density. Here the symbol sinn(x) stands for sinh(x) (if $\Omega_k>0$), sin(x) ($\Omega_k<0$) or just x ($\Omega_k=0$).

To compute the $\chi^2$ for the SNIa data, we follow \citet{2005PhRvD..72l3519N} to analytically marginalize over the nuisance parameter $H_0$.
\begin{equation}
\chi_{SN}^2 = A - 2\mu_0 B + \mu_0^2C,
\end{equation}
where 
\begin{equation}
\begin{aligned}
&A = \sum_{i=1}^{580} \frac{[\mu_{obs}(z_i)-\mu_{th}(z_i;\mu_0=0)]^2}{\sigma_i^2},\\
&B = \sum_{i=1}^{580} \frac{\mu_{obs}(z_i)-\mu_{th}(z_i;\mu_0=0)}{\sigma_i^2},\\
&C = \sum_{i=1}^{580} \frac{1}{\sigma_i^2}.
\end{aligned}
\end{equation}
$\sigma$ is the uncertainty in SNIa data. Eq. (5) has a minimum for $\mu_0 = B/C$ at
\begin{equation}
\tilde{\chi}_{SN}^2 = A - \frac{B^2}{C}.
\end{equation}
This equation is independent of $\mu_0$, so instead of $\chi_{SN}^2$ we will adopt $\tilde{\chi}_{SN}^2$ to compute the likelihood.

\subsection{Baryon Acoustic Oscillations (BAO)}

The competition between gravitational force and primordial relativistic plasma gives rise to acoustic oscillations which leaves its signature in every epoch of the universe. As standard rulers, BAOs provide another independent test for constraining the property of dark energy. 

\citet{2005ApJ...633..560E} first found a peak of this baryon acoustic oscillations in the 2-point correlation function at 100 $h^{-1}$ Mpc separation measured from the Sloan Digital Sky Survey (SDSS) Third Data Release (DR3) Luminous Red Galaxy (LRG) sample with effective redshift $z=0.35$. \citet{2010MNRAS.401.2148P} performed a power-spectrum analysis of the SDSS DR7 dataset, considering both the main and LRG samples, and measured the BAO signal at both $z=0.2$ and $z=0.35$. Recently, in the low-redshift universe the 6dF Galaxy Survey (6dFGS) team has reported a BAO detection at $z=0.1$ \citep{2011MNRAS.416.3017B}. Most recently, \citet{2011MNRAS.tmp.1598B} presented measurements of the BAO peak at redshifts $z=0.44,0.6$ and 0.73 in the galaxy correlation function of the final dataset of the WiggleZ Dark Energy Survey. They combined their WiggleZ BAO measurements with SDSS DR7 and 6dFGS datasets to give tight constraints on dark energy. In this work, \textbf{we} follow them to constrain different dark energy models using their combined BAO dataset. We highlight our usage of \textbf{this combined BAO dataset}, since there are altogether \textbf{six data points, which are more than} previous BAO data, and few have used this combined BAO datasets to constrain dark energy since the publication of the WiggleZ paper.

The data can be found in the above papers, but for completeness here we summarize the BAO measurements and the way to use them.

The $\chi^2$ for the WiggleZ BAO data is given by \citet{2011MNRAS.tmp.1598B},
\begin{equation}
\chi^2_{\scriptscriptstyle WiggleZ} = (\bar{A}_{obs}-\bar{A}_{th}) C_{\scriptscriptstyle WiggleZ}^{-1} (\bar{A}_{obs}-\bar{A}_{th})^T,
\end{equation}
where the data vector is $\bar{A}_{obs} = (0.474,0.442,0.424)$ for the effective redshift $z=0.44,0.6$ and 0.73. The corresponding theoretical value $\bar{A}_{th}$ denotes the acoustic parameter $A(z)$ introduced by \citet{2005ApJ...633..560E}:
\begin{equation}
A(z) = \frac{D_V(z)\sqrt{\Omega_{m}H_0^2}}{cz},
\end{equation}
and the distance scale $D_V$ is defined as
\begin{equation}
D_V(z)=\frac{1}{H_0}\left[(1+z)^2D_A(z)^2\frac{cz}{E(z)}\right]^{1/3},
\end{equation}
where $D_A(z)$ is the Hubble-free angular diameter distance which relates to the Hubble-free luminosity distance through $D_A(z)=D_L(z)/(1+z)^2$.
The inverse covariance $C_{\scriptscriptstyle WiggleZ}^{-1}$ is given by
\begin{equation}
C_{\scriptscriptstyle WiggleZ}^{-1} = \left(
\begin{array}{ccc}
1040.3 & -807.5 & 336.8\\
-807.5 & 3720.3 & -1551.9\\
336.8 & -1551.9 & 2914.9
\end{array}\right).
\end{equation}

Similarly, for the SDSS DR7 BAO distance measurements, the $\chi^2$ can be expressed as \citep{2010MNRAS.401.2148P}
\begin{equation}
\chi^2_{\scriptscriptstyle SDSS} = (\bar{d}_{obs}-\bar{d}_{th})C_{\scriptscriptstyle SDSS}^{-1}(\bar{d}_{obs}-\bar{d}_{th})^T,
\end{equation}
where $\bar{d}_{obs} = (0.1905,0.1097)$ is the datapoints at $z=0.2$ and $0.35$. $\bar{d}_{th}$ denotes the distance ratio 
\begin{equation}
d_z = \frac{r_s(z_d)}{D_V(z)}.
\end{equation}
Here, $r_s(z)$ is the comoving sound horizon,
\begin{equation}
 r_s(z) = c \int_z^\infty \frac{c_s(z')}{H(z')}dz',
 \end{equation}
where the sound speed $c_s(z) = 1/\sqrt{3(1+\bar{R_b}/(1+z)}$, with $\bar{R_b} = 31500 \Omega_{b}h^2(T_{CMB}/2.7\rm{K})^{-4}$ and $T_{CMB}$ = 2.726K.

The redshift $z_d$ at the baryon drag epoch is fitted with the formula proposed by      \citet{1998ApJ...496..605E},
\begin{equation}
z_d = \frac{1291(\Omega_{m}h^2)^{0.251}}{1+0.659(\Omega_{m}h^2)^{0.828}}[1+b_1(\Omega_b h^2)^{b_2}],
\end{equation}
where
\begin{equation}
\begin{aligned}
&b_1 = 0.313(\Omega_{m}h^2)^{-0.419}[1+0.607(\Omega_{m}h^2)^{0.674}], \\
&b_2 = 0.238(\Omega_{m}h^2)^{0.223}.
 \end{aligned}
\end{equation}

$C_{\scriptscriptstyle SDSS}^{-1}$ in Eq. (12) is the inverse covariance matrix for the SDSS data set given by
\begin{equation}
C_{\scriptscriptstyle SDSS}^{-1} = \left(
\begin{array}{cc}
30124 & -17227\\
-17227 & 86977
\end{array}\right).
\end{equation}

For the 6dFGS BAO data \citep{2011MNRAS.416.3017B}, there is only one data point at $z=0.106$, the $\chi^2$ is easy to compute:
\begin{equation}
\chi^2_{\scriptscriptstyle 6dFGS} = \left(\frac{d_z-0.336}{0.015}\right)^2.
\end{equation}

The total $\chi^2$ for all the BAO data sets thus can be written as
\begin{equation}
\chi^2_{BAO} = \chi^2_{\scriptscriptstyle WiggleZ} + \chi^2_{\scriptscriptstyle SDSS} + \chi^2_{\scriptscriptstyle 6dFGS}.
\end{equation}

\subsection{Cosmic Microwave Background (CMB)}
 Since the SNIa and BAO data contain information about the universe at relatively low redshifts, we will include the CMB information by \textbf{using the WMAP 7-yr data} \citep{2011ApJS..192...18K} to probe the entire expansion history up to the last scattering surface. The $\chi^2$ for the CMB data is constructed as
 \begin{equation}
 \chi^2_{CMB} = X^TC_{CMB}^{-1}X,
 \end{equation}
 where
 \begin{equation}
 X =\left(
 \begin{array}{c}
 l_A - 302.09 \\ 
 R - 1.725 \\
 z_* - 1091.3
 \end{array}\right).
 \end{equation}
 Here $l_A$ is the ``acoustic scale'' defined as
\begin{equation}
l_A = \frac{\pi d_L(z_*)}{(1+z)r_s(z_*)},
\end{equation}
where $d_L(z)=D_L(z)/H_0$ and the redshift of decoupling $z_*$ is given by \citet{1996ApJ...471..542H},
\begin{equation}
z_* = 1048[1+0.00124(\Omega_b h^2)^{-0.738}] [1+g_1(\Omega_{m}h^2)^{g_2}],
\end{equation}
\begin{equation}
g_1 = \frac{0.0783(\Omega_b h^2)^{-0.238}}{1+39.5(\Omega_b h^2)^{0.763}}, 
 g_2 = \frac{0.560}{1+21.1(\Omega_b h^2)^{1.81}},
\end{equation}

The ``shift parameter'' $R$ in Eq. (21) is defined as \citep{1997MNRAS.291L..33B}
\begin{equation}
R = \frac{\sqrt{\Omega_{m}}}{c(1+z_*)} D_L(z).
\end{equation}
$C_{CMB}^{-1}$ in Eq. (20) is the inverse covariance matrix,
\begin{equation}
C_{CMB}^{-1} = \left(
\begin{array}{ccc}
2.305 & 29.698 & -1.333\\
29.698 & 6825.270 & -113.180\\
-1.333 & -113.180 & 3.414
\end{array}\right).
\end{equation}

\subsection{Observational Hubble Data (OHD)}
In addition to the SNIa, BAO and CMB data, we also use the observational Hubble parameter as an observational technique. These data compose an independent dataset that can help break the parameter degeneracies, thus may also shed light on the cosmological models we aim to study. 

In this work, \textbf{we} adopt 11 data points from differential ages of old passive evolving galaxies \citep{2010JCAP...02..008S}, the $\chi^2$ value for these OHD can be expressed as
\begin{equation}
\chi^2_{OHD} = \sum_{i=1}^{11} \frac{[H_{th}(z_i)-H_{obs}(z_i)]^2}{\sigma_i^2},
\end{equation}
where $\sigma_i$ is the 1$\sigma$ error in the OHD data with $z_i$ ranging from 0.1 to 1.75.

\section{cosmological models and constraining results}
In the following, we study eight popular cosmological models discussed in the literature. The models with their parameters and the abbreviations we use are listed in Table 1. We examine them through the expansion history of the universe to see whether they are consistent with current data at the background level. The model parameters are determined through the minimum $\chi^2$ fitting by using the Markov Chain Monte Carlo (MCMC) method. Our MCMC code is based on the publicly available CosmoMC package \citep{2002PhRvD..66j3511L}.

It should be stressed here that unlike most other work on dark energy model constraints, we do not assume a spacially flat universe as a prior in this paper, although recent studies showed that the universe is nearly flat \citep{2011ApJS..192...18K}. When we constrain the properties of dark energy, the parameters such as the equation of state $w$, are always degenerate with the curvature density $\Omega_k$. It has already been shown that ignoring $\Omega_k$ will induce large errors on the reconstructed dark energy parameter (e.g. $w$). If the true geometry is not flat and with the wrong flatness assumption, one will erroneously conclude a wrong behavior of dark energy even if the curvature term is very small \citep{2007JCAP...08..011C,2007PhLB..648....8Z,2008JCAP...12..008V}. So in our work, instead of assuming a flat universe, we will include $\Omega_{k}$ as a free parameter in different cosmological models.

\begin{table}
\caption{Summary of cosmological models}
\begin{tabular}{ccc}
\hline \hline
Model & Abbreviation & Parameters\\
\hline
Cosmological constant & $\Lambda$CDM & $\Omega_k$, $\Omega_m$ \\
Constant $w$ & $w$CDM & $\Omega_k$, $\Omega_m$, $w$\\
Varying $w$ (CPL) & CPL & $\Omega_k$, $\Omega_m$, $w_0$, $w_a$\\
Generalized Chaplygin Gas & GCG & $\Omega_k$, $A_s$, $\alpha$ \\
Dvali-Gabadadze-Porrati & DGP & $\Omega_k$, $\Omega_m$ \\
Modified Polytropic Cardassian & MPC & $\Omega_k$, $\Omega_m$, $q$, $n$\\
Interacting Dark Energy & IDE & $\Omega_k$, $\Omega_m$, $w_x$, $\delta$ \\
Early Dark Energy & EDE & $\Omega_k$, $\Omega_m$, $\Omega_e$, $w_0$\\
\hline
\end{tabular}
Note: The Hubble constant $H_0$ in the fit is not deemed as a model parameter, but we include it in the number of degrees of freedom and in $k$ when calculating the AIC and BIC.
\end{table}

\subsection{Cosmological constant model}
The cosmological constant $\Lambda$ was originally introduced by \citet{A.E.1917} to achieve a static universe but later abandoned by Einstein after Hubble's discovery of the expansion of the universe. Ironically, after 1998 the cosmological constant revived again as a form of dark energy responsible for the late-time acceleration of the universe. The cosmological constant plus cold dark matter (CDM) is usually called the  $\Lambda$CDM model, and in this model the dark energy equation of state $w=-1$ at all times. The Friedmann equation in this case is
\begin{equation}
H^2(z)/ H_0^2=\Omega_k(1+z)^2+\Omega_r(1+z)^4+\Omega_m(1+z)^3+(1-\Omega_k-\Omega_r-\Omega_m),
\end{equation}
where the radiation density parameter $\Omega_r$ is given by $\Omega_r=\Omega_\gamma(1+0.2271N_{eff})$ with $\Omega_\gamma=2.469\times10^{-5}h^{-2}$ and the effective number of neutrino species $N_{eff}=3.04$ \citep{2011ApJS..192...18K}. We caution that in many papers the $\Omega_r$ term is usually neglected. While this is reasonable for SNIa analysis where the redshift is very small, for high redshift especially at CMB epoch this term is dominated. When calculating the sound horizon $r_s$ for CMB and BAO analysis, ignoring this radiation term will induce large errors on the results, so it should better be included. The last term in the equation represents the energy density of the cosmological constant.

This simple model has only two parameters $\Omega_k$ and $\Omega_m$. Our global fitting from all the four data sets gives the best-fit values with 1 $\sigma$ errors:
\begin{equation}
\Omega_k = -0.0024 \pm 0.0056, ~~\Omega_m =  0.291 \pm  0.014.
\end{equation}

Our results are consistent with the latest results of the WiggleZ BAO paper \citep{2011MNRAS.tmp.1598B}. Fig. 1 shows the constraint from each of the SNIa, CMB and BAO data sets and the joint constraint from all the four data sets. We do not separately give the constraint from OHD data since currently it is not as stringent as the first three probes, but we include it in the combined results. It can be seen that although the contour of each single data set is quite broad, their combined constraint is quite stringent and this reminds us of the power of joint analysis from different independent data sets. A flat universe is quite favored by current data within 1$\sigma$ confidence level.

\subsection{Constant $w$ model}
The simplest extension to the $\Lambda$CDM model is to assume that the dark energy equation of state $w$ does not precisely equals $-1$, but a constant to be fitted with data. In this model, the Friedmann equation is
\begin{multline}
H^2(z)/ H_0^2=\Omega_k(1+z)^2+\Omega_r(1+z)^4+\Omega_m(1+z)^3\\
+ (1-\Omega_k-\Omega_r-\Omega_m)(1+z)^{3(1+w)}.
\end{multline}

There are three parameters in this model: ~$\Omega_k$, $\Omega_m$, and $w$. The best-fit values using all the data sets are
\begin{equation}
\Omega_k =  -0.0012 \pm  0.0064, ~~\Omega_m =  0.292 \pm 0.015, \\
w =  -0.990 \pm  0.041,
\end{equation}
also in agreement with \citet{2011MNRAS.tmp.1598B}.

We plot the contour of $\Omega_m$ and $w$ after marginalizing over $\Omega_k$ and $H_0$ in Fig. 2. This model also gives a good fit to different data sets. The combined result shows a clear preference around the cosmological constant model ($w=-1$ within 1$\sigma$ confidence level).

\subsection{Chevallier-Polarski-Linder model}
There is no prior reason to expect $w$ to be $-1$ or a constant, and if $w$ varies with time, the Friedmann equation is modified as
\begin{multline}
H^2(z)/ H_0^2=\Omega_k(1+z)^2+\Omega_r(1+z)^4+\Omega_m(1+z)^3\\
+ (1-\Omega_k-\Omega_r-\Omega_m)~\exp \left(3\int^z_0 \frac{1+w(z')}{1+z'}dz'\right).
\end{multline}

Many function forms of $w$ evolving with redshift have been proposed so far (see, e.g. \citealt{2007IJMPD..16.1581J}). Among the various parametrizations of dark energy equation of state $w$, the one developed by \citet{2001IJMPD..10..213C} and \citet{2003PhRvL..90i1301L} turns out to be an excellent approximation to a wide variety of dark energy models, and this CPL (Chevalier-Polarski-Linder) model is the most commonly used function form to study the time dependence of $w$. The equation of state in this model is
\begin{equation}
w(z) = w_0 + w_a\frac{z}{1+z},
\end{equation}

So insert Eq. (33) into Eq. (32), we get the Friedmann equation for this CPL model:
\begin{multline}
H^2(z)/ H_0^2=\Omega_k(1+z)^2+\Omega_r(1+z)^4+\Omega_m(1+z)^3\\
+ (1-\Omega_k-\Omega_r-\Omega_m)~(1+z)^{3(1+w_0+w_a)} \exp (\frac{-3 w_a z}{1+z}).
\end{multline}

There are four parameters in this model:~$\Omega_k$, $\Omega_m$, $w_0$ and $w_a$. Our best-fit values for these parameters are
\begin{equation}
\begin{aligned}
& \Omega_k =  0.00027 ^{+0.0034}_{-0.0051},~\Omega_m = 0.293 \pm 0.016,\\
& w_0 = -0.966^{+0.088}_{-0.105}, ~w_a = 0.202^{+1.030}_{-1.053}.
\end{aligned}
\end{equation}

Fig. 3 shows the contours of $w_0$ and $w_a$ for the CPL model after marginalized over other parameters. Obviously, $w_a$ is weakly constrained by current data. This is partially due to the degeneracy between the curvature and the equation of state. If we set $\Omega_k=0$ in the fit as most work did, the constraints would be more stringent, especially for a single data set. However, as explained earlier, we don't assume a flat prior in the fitting procedure. We see once again that it is consistent with the $\Lambda$CDM model for $w_0=-1$ and $w_a=0$. Our results are in agreement with \citet{2011MNRAS.tmp.1598B} although they assumed a flat universe in their fit.

\subsection{Modified Polytropic Cardassian expansion}
The Cardassian expansion model was first proposed in \citet{2002PhLB..540....1F} which modifies the Friedmann equation to allow for an acceleration in a matter-dominated universe. The motivation for this modification could be the embedding of our observable universe living as a 3-dimensional brane in a higher
dimensional universe. The original form of the Cardassian model can be written as
\begin{equation}
H^2(z) = \frac{8\pi G}{3}\rho_m + B \rho_m^n,
\end{equation}
where $B$ is a constant and $n$ is a dimensionless parameter.

This power law form is equivalent to the constant $w$ model (Sec. 4.2) for $w=n-1$, so there is no need to additionally fit this model. Here we consider a modified polytropic Cardassian model proposed by \citet{2003ApJ...594...25W}, and in addition, we also include the curvature and radiation term :
\begin{multline}
H^2(z)/ H_0^2 =\Omega_k(1+z)^2+\Omega_r(1+z)^4\\
+\Omega_m(1+z)^3\left[1+\left((\frac{1-\Omega_k-\Omega_r}{\Omega_m})^q-1\right)(1+z)^{3q(n-1)}\right]^{\frac{1}{q}}.
\end{multline}

The above equation reduces to the $\Lambda$CDM one for $q=1$ and $n=0$. Our joint constraints give the best-fit parameters as follows:
\begin{equation}
\begin{aligned}
&\Omega_k = 0.0022 \pm 0.0025,~\Omega_m = 0.280 \pm 0.006,\\
&q = 0.897^{+0.152}_{-0.468},~n = -0.648^{+0.856}_{-1.106}.
\end{aligned}
\end{equation}

The constraints on the parameter $q$ and $n$ is very weak from current data. Fig. 4 displays the marginalized contours of $q$ and $n$. It can be seen that it is still consistent with the $\Lambda$CDM model in 1$\sigma$ level. 

\subsection{Dvali-Gabadadze-Porrati model}
The Dvali-Gabadadze-Porrati (DGP) model is a popular model which modifies the gravity to allow for cosmic acceleration without dark energy \citep{2000PhLB..485..208D}. This model may arise from the brane world theory in which gravity leaks out into the bulk at large scales. The Friedmann equation is modified as
\begin{multline}
H^2(z)/ H_0^2 =\Omega_k(1+z)^2+\Omega_r(1+z)^4\\
+\left(\sqrt{\Omega_m(1+z)^3+\Omega_{r_c}}+\sqrt{\Omega_{r_c}}\right)^2,
\end{multline}
where $r_c$ is the length scale beyond which gravity leaks out into the bulk, and $\Omega_{r_c}=1/(4r_c^2H_0^2)$. Setting $z=0$ in Eq. (39), we get the normalization condition
\begin{equation}
 \Omega_{r_c}=\frac{(1-\Omega_m-\Omega_r-\Omega_k)^2}{4(1-\Omega_k-\Omega_r)}.
\end{equation}

The DGP model has the same number of parameters as the $\Lambda$CDM model. The marginalized best-fit parameters are as follows:
\begin{equation}
\Omega_k = 0.020 \pm 0.006, ~\Omega_m = 0.305 \pm 0.015.
\end{equation}

We can see that although the matter density is consistent with that of the $\Lambda$CDM one, the curvature term is much larger than that in other models. This feature is also noticed by \citet{2005ApJ...620....7Z} and \citet{2006ApJ...646....1G} who got a non-flat universe for the DGP model at high confidence level. In Fig. 5, it can be seen that the three observational probes strongly disagree -- the areas of intersection of any pair are distinct from other pairs.  CMB data prefer a positive $\Omega_k$ while SN and BAO data are in support of negative $\Omega_k$. \citet{2009ApJ...695..391R} and \citet{2007ApJ...666..716D} also noticed this signal. It may imply that this DGP model is strongly disfavored by current data. This can be further quantified by the model selection statistics to be shown in Sec. 5. 

It should be mentioned that the DGP model could perform better when using only SNIa data. For example, using MLCS2k2 light curve fitter for SDSS-II supernova data, \citet{2009ApJ...703.1374S} found that the DGP model perform better than the $\Lambda$CDM model under the information criteria. Also, recently it was noticed that the Supernova Legacy Survey (SNLS) three years data analyzed with SALT2 fitters alone prefer the DGP model over others \citep{2012ApJ...744..176L}. However, when combining with BAO and CMB data, things have changed, and the concordance $\Lambda$CDM model became favored. This is not surprising, since current SNIa data are mainly confined by systematic errors rather than statistical errors, so it would be better to combine SNIa data set with other probes (\textbf{BAO, CMB, etc.}) to constrain cosmological models to avoid any potential bias that may be caused by the systematics of SNIa.

\subsection{Interacting Dark Energy model}
The fact that the energy density of dark energy is the same order as that of dark matter in the present universe suggests that there may be some relations between them. This may rise from the an interaction between a scalar field (e.g. quintessence field) and dark matter. Such motivation may help alleviate the coincidence problem. A popular approach to study this interaction is to introduce a coupling term on the right hand side of the continuity equations \citep{2001PhRvL..87n1302D,2005JCAP...03..002C,2007PhRvD..76b3508G,2009PhRvD..79f3518C}:
\begin{equation}
\begin{aligned}
&\dot{\rho}_m + 3H \rho_m = +\Gamma \rho_m,\\
&\dot{\rho}_x + 3H (1+w_x)\rho_x = -\Gamma \rho_m,
\end{aligned}
\end{equation}
where $\Gamma$ characterizes the strength of the interacting, $\rho_m$ is the matter density and $\rho_x$ is the dark energy density with $w_x$ the equation of state. In order to place observational constraints on the coupling term $\Gamma$, it is convenient to express $\Gamma$ in term of the Hubble parameter $H$ 
\begin{equation}
\Gamma = \delta H,
\end{equation}
where $\delta$ is a dimensionless coupling term. Note that a positive $\delta$ corresponds to a transfer of energy from dark energy to dark matter, whereas for a negative $\delta$ the energy transfer is opposite.

It is obvious the expansion history will depend on the parameter $\delta$, and thus we are interested in placing observational constraints on it. For simplicity, here we assume $\delta$ to be a constant. In more general case $\delta$ may be varying, and there have already been a lot of work on this varying case. In this paper since we mainly focus on the model comparison, studying a constant coupling is enough for our purpose.

For a constant $\delta$, solving Eq. (42) with Eq. (43), the Friedmann equation becomes
\begin{multline}
H^2(z)/ H_0^2 =\Omega_k(1+z)^2+\Omega_r(1+z)^4+(1-\Omega_m-\Omega_k-\Omega_r)(1+z)^{3(1+w_x)}\\
+\frac{\Omega_m}{\delta+3w_x}\left[\delta(1+z)^{3(1+w_x)}+3w_x(1+z)^{3-\delta}\right].
\end{multline}

This model has four parameters: $\Omega_k$, $\Omega_m$, $w_x$ and $\delta$. The concordance $\Lambda$CDM model is recovered for $\delta=0$ and $w_x=-1$. Our global fitting gives the following best-fit values
\begin{equation}
\begin{aligned}
&\Omega_k = 0.0007 \pm 0.0032, ~\Omega_m = 0.292 \pm 0.007,\\
&\delta = -0.0043 \pm 0.0066, ~w_x = -1.001 \pm 0.087.
\end{aligned}
\end{equation}

Fig. 6 shows the case for this interacting dark energy model. It is noticed that the SNIa and BAO data sets give quite weak constraints on the parameter space, comparing to the CMB data set. This is not strange since the SNIa and BAO data are located in low redshifts, and we can see from Eq. (44) that when $z\ll1$ the $\delta+3w_x$ term just cancels out, so the corresponding information about $\delta$ lost. This tells us how important it is to include other high redshift data. The CMB and OHD data are appropriate for this purpose. Our results show that the $\Lambda$CDM model still remains a good fit to the data (at least within 2 $\sigma$ level), but a negative coupling ($\delta<0$), i.e., the energy transfers from dark matter to dark energy, is slightly favored. Also in this case the equation of state of dark energy $w_x$ prefers a phantom case $w_x<-1$. This result is consistent with that obtained by other authors \citep{2007PhRvD..76b3508G,2010MNRAS.402.2355V,2011MNRAS.416.1099C}.

\subsection{Generalized Chaplygin Gas model}
Despite of their quite different properties in the equation of state and clustering, the temptation to unify dark energy and dark matter in a single entity has occured to many researchers from the beginning. To realize it, a natural and simple way is to introduce a perfect background fluid. The Chaplygin gas model is just a typical example. 

The original Chaplygin gas model was proposed by \citet{2001PhLB..511..265K}. In this model, the pressure $P$ of the fluid is related to its energy density $\rho$ through $P=-A/\rho$ where $A$ is a positive constant. In a more general case, one may consider a generalized chaplygin gas model (GCG) given by \citet{2002PhRvD..66d3507B}
\begin{equation}
P = -A \rho^\alpha.
\end{equation}
Consider the energy conservation in the framework of Friedmann-Robertson-Walker (FRW) metric, we obtain the following solution
\begin{equation}
\rho(a) = \rho_0\left[A_s+\frac{1-A_s}{a^{3(1+\alpha)}}\right]^\frac{1}{1+\alpha},
\end{equation}
where $A_s=A/\rho_0^{1+\alpha}$, and $\rho_0$ is the present energy density of the GCG. One finds the 
intriguing feature that the energy density of this GCG acts like dust matter in the early time and behaves as a cosmological constant at late epoch. So the GCG model can account for both dark matter and dark energy at the background level. The Friedmann equation for this model can be written as 
\begin{multline}
H^2(z)/ H_0^2 =\Omega_k(1+z)^2+\Omega_r(1+z)^4 + \Omega_b(1+z)^3 \\
+ (1-\Omega_k-\Omega_r-\Omega_b)[A_s+(1-A_s)(1+z)^{3(1+\alpha)}]^\frac{1}{1+\alpha},
\end{multline}
where $\Omega_b$ is the present density parameter of the baryonic matter. We adopt $\Omega_b=0.0451$ according to the WMAP 7-year results \citep{2011ApJS..192...18K}. The effective total matter density can be expressed as $\Omega_m=\Omega_b+(1-\Omega_b-\Omega_k-\Omega_r)(1-A_s)^{1/(1+\alpha)}$. Note that the concordance $\Lambda$CDM model is recovered by $\alpha=0$, thus $\Omega_m=1-\Omega_k -\Omega_r-A_s(1-\Omega_b-\Omega_k-\Omega_r)$.

There are three parameters in this model. The fitting results are:
\begin{equation}
\Omega_k = 0.0004 \pm 0.0032, ~A_s = 0.733 \pm 0.025,~ \alpha = -0.011 \pm 0.140.
\end{equation}

The GCG model provides a good fit to the data. Fig. 7 shows the contours for the two parameters $A_s$ and $\alpha$ in this GCG model. It can be seen that the $\Lambda$CDM model ($\alpha=0$) falls well within 1$\sigma$ level, and the original Chaplygin gas model ($\alpha=1$) is ruled out at more than 2$\sigma$ confidence level. This is in agreement with the results of \citet{2011A&A...527A..11L} and \citet{2007ApJ...658..663W}. 

\subsection{Early Dark Energy scenario}
One of the differences between dynamical dark energy and the cosmological constant is that the energy density of the former may be non-negligible even at very high redshift (e.g. around recombination, or earlier). The existence of the so-called ``tracker'' field \citep{1999PhRvD..59l3504S} is important to alleviate the cosmological constant problem. The tracker fields correspond to attractor-like solutions in which the field energy density tracks the background fluid density for a wide range of initial conditions. 
These models can be motivated by dilatation symmetry in particle physics and string theory \citep{1988NuPhB.302..668W}.

As a specific model of such early dark energy scenario, here we consider a commonly used form with the dark energy density expressed as \citep{2006JCAP...06..026D}
\begin{equation}
\Omega_{DE}(z) = \frac{\Omega_{DE}^{0}-\Omega_e[1-(1+z)^{3w_0}]}{\Omega_{DE}^0+\Omega_m(1+z)^{-3w_0}}+\Omega_e[1-(1+z)^{3w_0}],
\end{equation}
where $\Omega_{DE}^{0}$ is the present dark energy density, $\Omega_e$ is the asymptotic early dark energy density and $w_0$ is the present dark energy equation of state. This equation is based on simple considerations as depicted in \citet{2006JCAP...06..026D} and \citet{2001PhRvD..64l3520D}. The early dark energy behavior is included in the $\Omega_e$ term. The $-3w_0$ term, motivated by the relation $\Omega_{DE}(z)/\Omega_m(z)\propto (1+z)^{3w}$, allows the deviation from the $\Lambda$CDM model. 

Eq. (50) assumes a spacially flat universe, and since in this work we do not assume flatness from the beginning, we would like to slightly modify that equation to include a contribution from curvature,
\begin{multline}
 \Omega_{DE}(z) = \left(\Omega_{DE}^0-\Omega_e[1-(1+z)^{3w_0}]\right)/\\
 [\Omega_{DE}^0+\Omega_m(1+z)^{-3w_0}+\Omega_r(1+z)^{-3w_0+1}+\Omega_k(1+z)^{-3w_0-1}]\\
 +\Omega_e[1-(1+z)^{3w_0}],
\end{multline}
where $\Omega_{DE}^{0}=1-\Omega_m-\Omega_r-\Omega_k$.

In this case, the Friedmann equation can be expressed as
\begin{equation}
H^2(z)/ H_0^2 = \frac{\Omega_m(1+z)^3+\Omega_r(1+z)^4+\Omega_k(1+z)^2}{1-\Omega_{DE}(z)}.
\end{equation}

This model also has four parameters. The best-fit values are as follows
\begin{equation}
\begin{aligned}
&\Omega_k = 0.0042 \pm 0.0069, ~\Omega_m = 0.291 \pm 0.007,\\
&\Omega_e = 0.026^{+0.007}_{-0.026}, ~w_0 = -1.039 \pm 0.097.
\end{aligned}
\end{equation}

As we can see from Fig. 8, the $\Lambda$CDM model ($\Omega_e=0, w_0=-1$) is still favored within 2$\sigma$ level. However, the early dark energy component is not totally excluded from current data. Since the SNIa and BAO data sets are low-redshift ones, they cannot give effective constraints on the early dark energy density $\Omega_e$, so the most stringent constraint comes from the CMB data set. The results are consistent with those of \citet{2009ApJ...695..391R}, \citet{2011PhRvD..83l3504C} and \citet{2011arXiv1110.5328R}. 

\section{Model Comparison}
In this section, we compare the above different models by using the model selection statistics. Table 2 gives a summary of the IC results. It is easy to see that the concordance $\Lambda$CDM model has the lowest ICs, so the $\Delta$AIC and $\Delta$BIC are all calculated with respect to the $\Lambda$CDM model.

Given current data sets, the $\Lambda$CDM model is clearly preferred by these model selection tests. Following it are a series of models that give comparably good fits but have more parameters. According to their ICs, we can roughly rank these models to four groups: 1. positive against (GCG,$w$CDM), 2. strong against (IDE,EDE), 3. very strong against (MPC,CPL), 4. essentially no support (DGP) from current data. The GCG and $w$CDM model fit the data well, maybe due to their fewer parameters and that they can easily reduce to the $\Lambda$CDM model. The IDE and EDE models are punished by the ICs mainly because \textbf{they have more parameters}. The constraints on the MPC and CPL models are very weak, and they are also penalized by their large number of parameters. We see that the DGP model is so strongly disfavored by the data as its $\Delta$ICs have much larger values than others. Its goodness of fit is also much smaller than others. So we can say, at least at the background level, the DGP model can be excluded by current joint data sets at high significance from a model selection point of view.

To see more clearly how to realize cosmic acceleration from these models, we plot the deceleration parameter $q$ in Fig. 9. The deceleration parameter $q$, defined as $q=-\ddot{a}a/\dot{a}^2$, can be calculated by
\begin{equation}
q=-1+\frac{1+z}{H(z)}~\frac{dH(z)}{dz}.
\end{equation}

As expected, these models all give negative $q$ at late times, and positive $q$ at earlier epoch, meaning that the expansion of the universe slowed down in the past and speeded up recently. Phenomenologically, there is a transition redshift $z_t$ between the two epochs, and we also give it in the figure. We can see from Fig. 9 that much due to their complexity, the constraints on the CPL and MPC models are very weak, as the contours of their parameters. Although the constraint on the DGP model is quite tight, it gives the transition redshift $z_t=0.45$, much smaller than the other models, suggesting a strong distinction of the DGP model from other models. Given the bad behavior of the DGP model from the model selection techniques discussed earlier, this smaller transition redshift, too, may suggest that the DGP model is disfavored by current data. The concordance $\Lambda$CDM model remains the best fit in the figure.

\section{Discussions and Conclusion}
In this work, we have studied a number of different cosmological models in light of the latest observational data. The data we used include the newly published Union2.1 supernovae compilation and the WiggleZ BAO measurements, together with the WMAP 7-year distance priors and the observational Hubble data. By using these data sets, we obtained the best-fit parameters for different models. We use the information criteria including the AIC and the BIC, to compare different models and to see which is the most favored one by current data. These ICs tend to favor models that give a good fit with fewer parameters. Unlike many previous work did, we do not assume a spacially flat universe in our work, instead, we treat the spacial curvature $\Omega_k$ as a free parameter in the fitting procedure. 

Using the AIC and BIC for model comparison, it is found that the concordance $\Lambda$CDM model remains the best one to explain current data. The generalized chaplygin gas model and the constant $w$ model also give good fits to the data. The interacting dark energy model,  the early dark energy scenario, the Chevallier-Polarski-Linder model and the modified polytropic cardassian model are all punished by their large number of parameters, thus are not favored by the ICs. The DGP model gives the worst fit, although it has the same number of parameters as the $\Lambda$CDM model. Its AIC and BIC are much larger than other models, with a bad goodness-of-fit. \textbf{Meanwhile, the curvature density parameter $\Omega_k$ is quite near zero for all models except the DGP model. In the DGP model the different contours from the different observational data sets strongly disagree--SNIa and BAO prefer negative value of $\Omega_k$ whereas the CMB prefers positive $\Omega_k$, and so the joint constraint on the value of $\Omega_k$ is much larger than in other models.} 

We also showed the deceleration parameter $q$ for different models, finding that all models indicate a late time cosmic acceleration consistent with observations. However, the transition redshift $z_t$ for the DGP model is much smaller than that in other models. This may reflect the fact that the DGP model can not reduce to the concordance $\Lambda$CDM model for any value of its parameters.   

In brief, given current data sets, the $\Lambda$CDM model remains the best one from a model-comparison point of view, followed by those that can reduce to it. Those who can not reduce to the concordance model fit the data quite \textbf{badly}, especially for the DGP model. In spite of its observational success, due to the theoretical considerations, we can not yet say that this $\Lambda$CDM model truly describes our universe. For the time being, we can at most conclude that this model fit the current data best among various models. With more and more precise data available in the future, it is excepted that we will finally be able to identify the nature of cosmic acceleration.
 
 \section*{Acknowledgments}
 K. Shi thanks Shi Qi for helpful comments and discussions.  The authors are also grateful to the anonymous referee for helpful comments and suggestions. This work was supported by the National Natural Science Foundation of China (grant nos. 10973039 and 11033002), the National Basic Research Program of China (973 program, grant no. 2009CB824800).
 
\begin{table}
\caption{\textbf{Summary of the information criteria results}}
\begin{tabular}{ccccc}
\hline \hline
Model & $\chi^2$/dof & GoF($\%$) & $\Delta$AIC & $\Delta$BIC\\
\hline
$\Lambda$CDM & 555.98/597 & 88.42 & 0.00 & 0.00\\
GCG & 555.64/596 & 88.04 & 1.66 & 6.06\\
$w$CDM & 555.96/596 & 87.85 & 1.98 & 6.38\\
IDE & 555.02/595 & 87.83 & 3.04 & 11.83\\
EDE & 555.06/595 & 87.81 & 3.08 & 11.87\\
MPC & 555.56/595 & 87.50 & 3.58 & 12.37\\
CPL & 555.94/595 & 87.26 & 3.96 & 12.75\\
DGP & 567.98/597 & 76.79 & 13.01 & 13.01\\
\hline
\end{tabular}

Note: The $\Lambda$CDM model is preferred by both the AIC and the BIC. Thus the $\Delta$AIC and $\Delta$BIC values for all other models are measured with respect to the $\Lambda$CDM model. The models are listed in order of increasing $\Delta$AIC. The goodness of fit (GoF) approximates the probability of finding a worse fit to the data. 
\end{table}

\begin{figure*}
\includegraphics[width=7in]{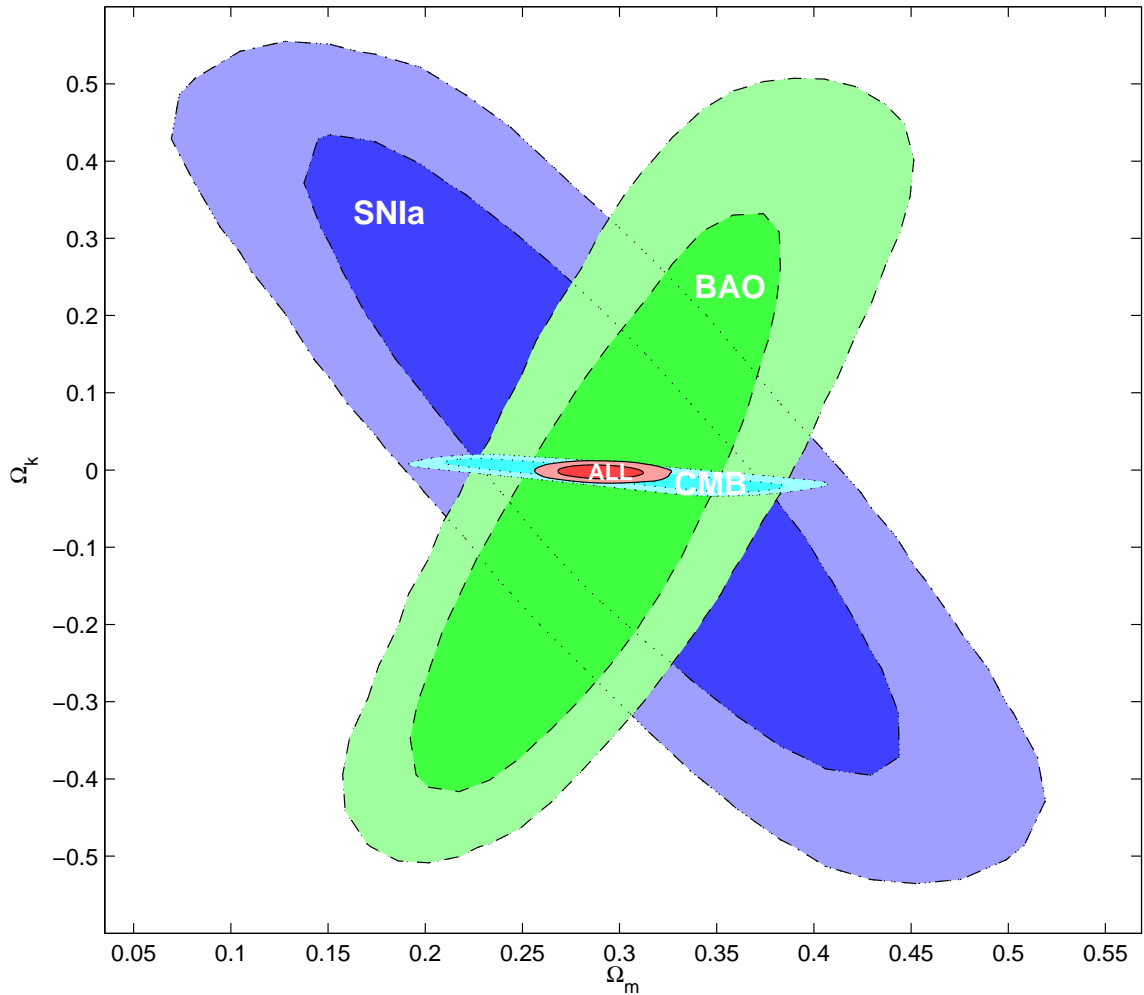}
\caption{The marginalized 1$\sigma$ and 2$\sigma$ contours of the $\Lambda$CDM model parameters $\Omega_k$ and $\Omega_m$, derived from different data sets. ``ALL'' denotes the joint constraint including all the four data sets.}
\end{figure*}

\begin{figure*}
\includegraphics[width=7in]{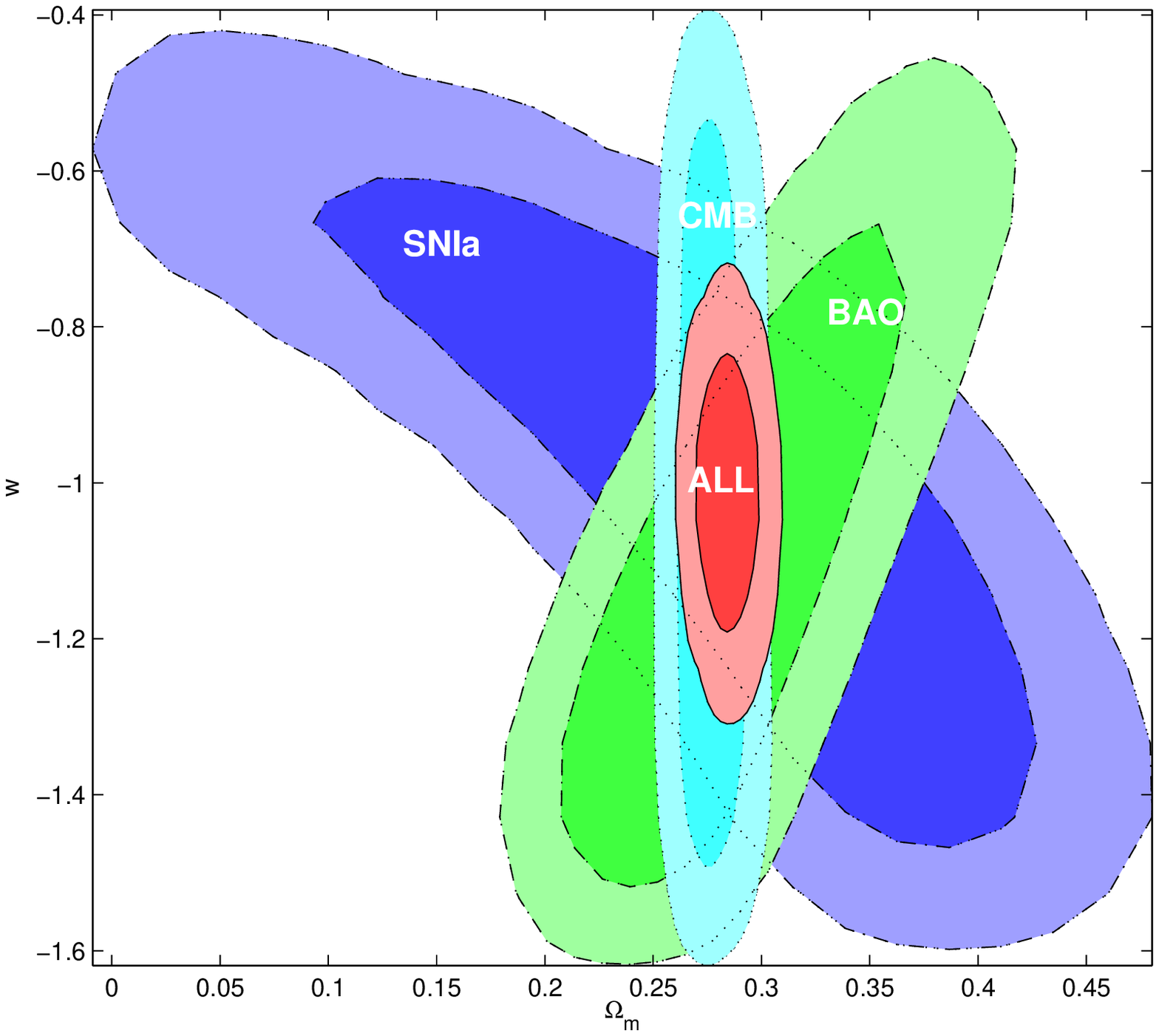}
\caption{The marginalized 1$\sigma$ and 2$\sigma$ contours of the $w$CDM model parameters $\Omega_m$ and $w$, derived from different data sets. ``ALL'' denotes the joint constraint including all the four data sets.}
\end{figure*}

\begin{figure*}
\includegraphics[width=7in]{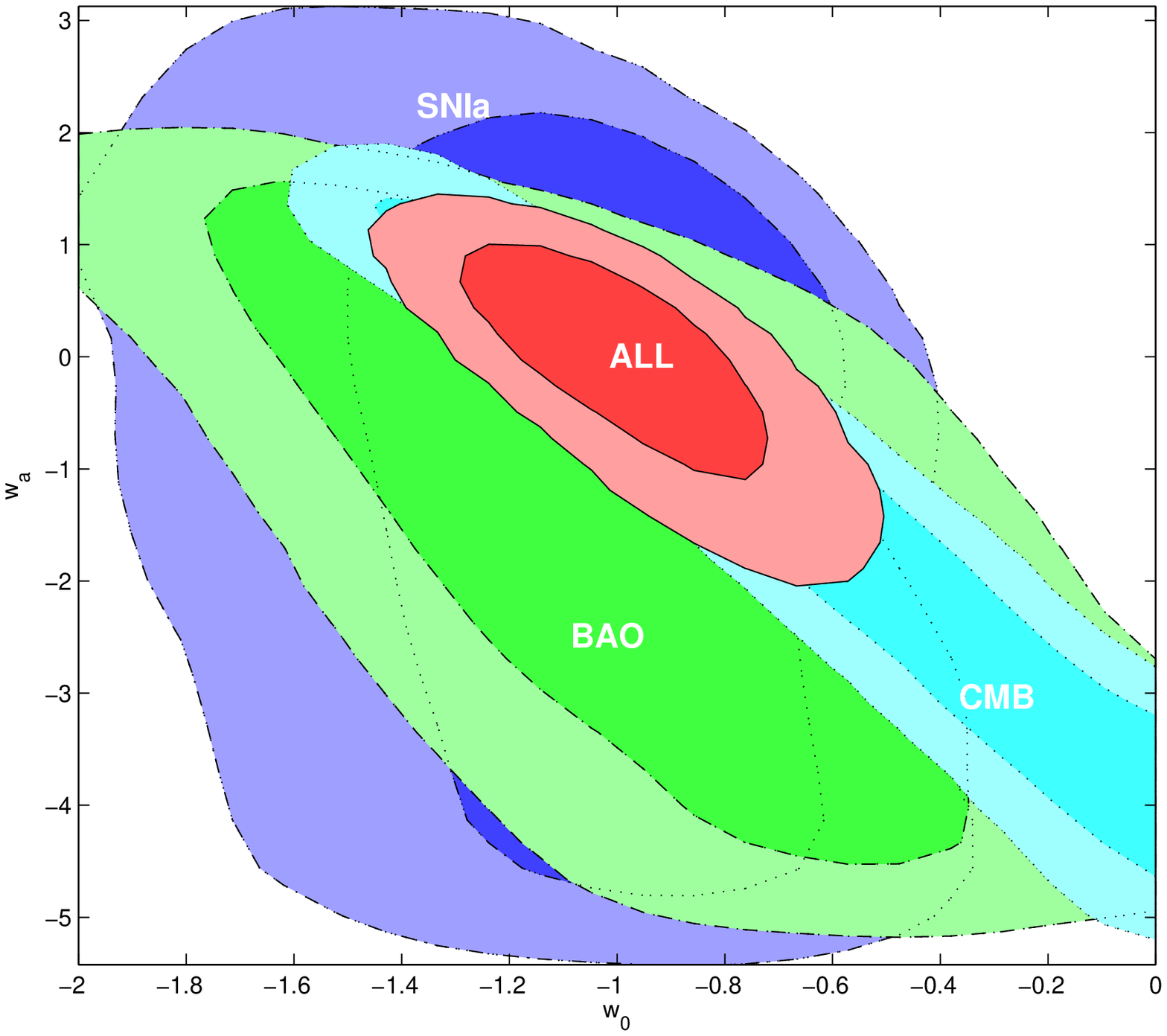}
\caption{The marginalized 1$\sigma$ and 2$\sigma$ contours of the CPL model parameters $w_0$ and $w_a$, derived from different data sets. ``ALL'' denotes the joint constraint including all the four data sets.}
\end{figure*}

\begin{figure*}
\includegraphics[width=7in]{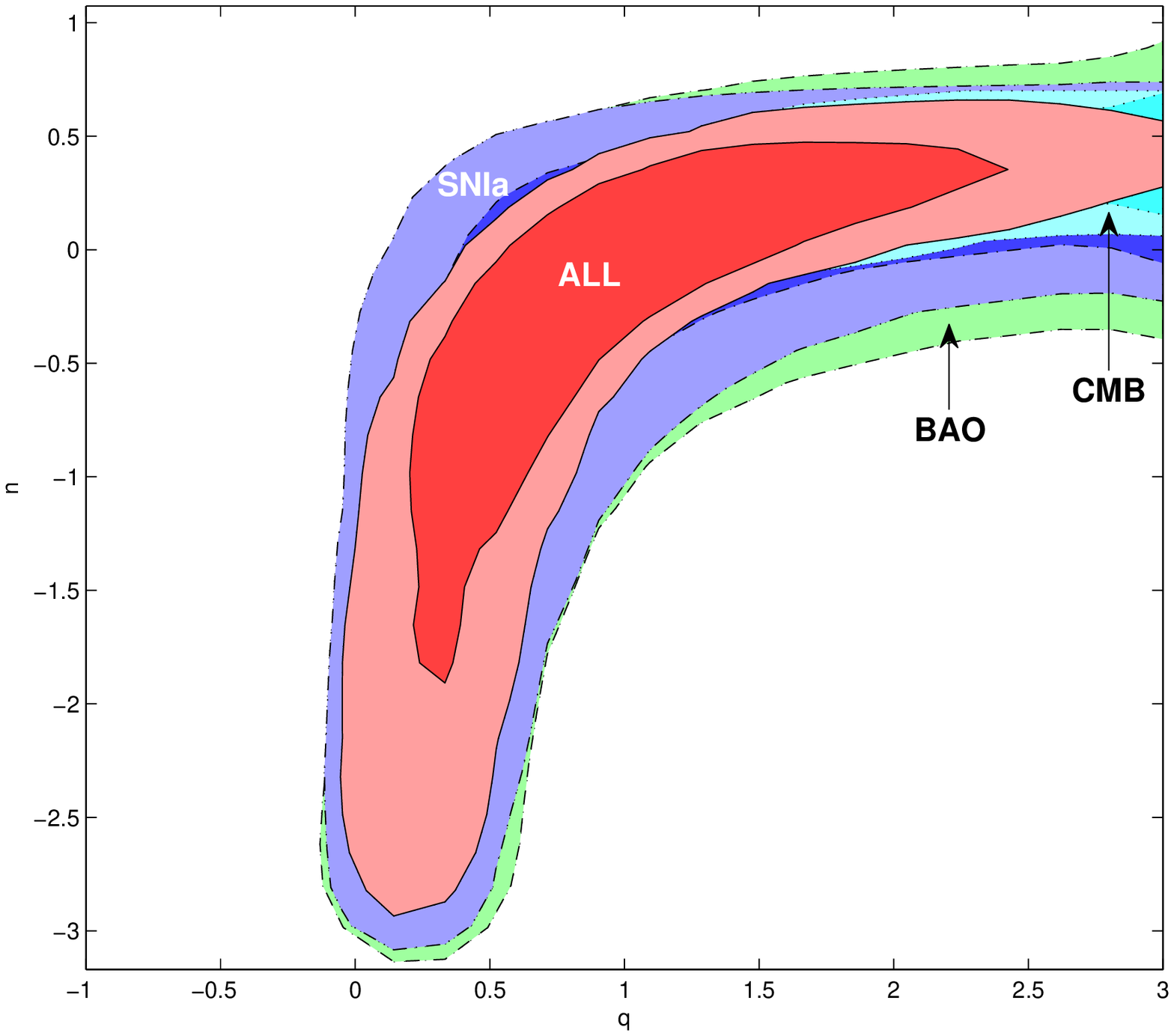}
\caption{The marginalized 1$\sigma$ and 2$\sigma$ contours of the MPC model parameters $q$ and $n$, derived from different data sets. ``ALL'' denotes the joint constraint including all the four data sets.}
\end{figure*}

\begin{figure*}
\includegraphics[width=7in]{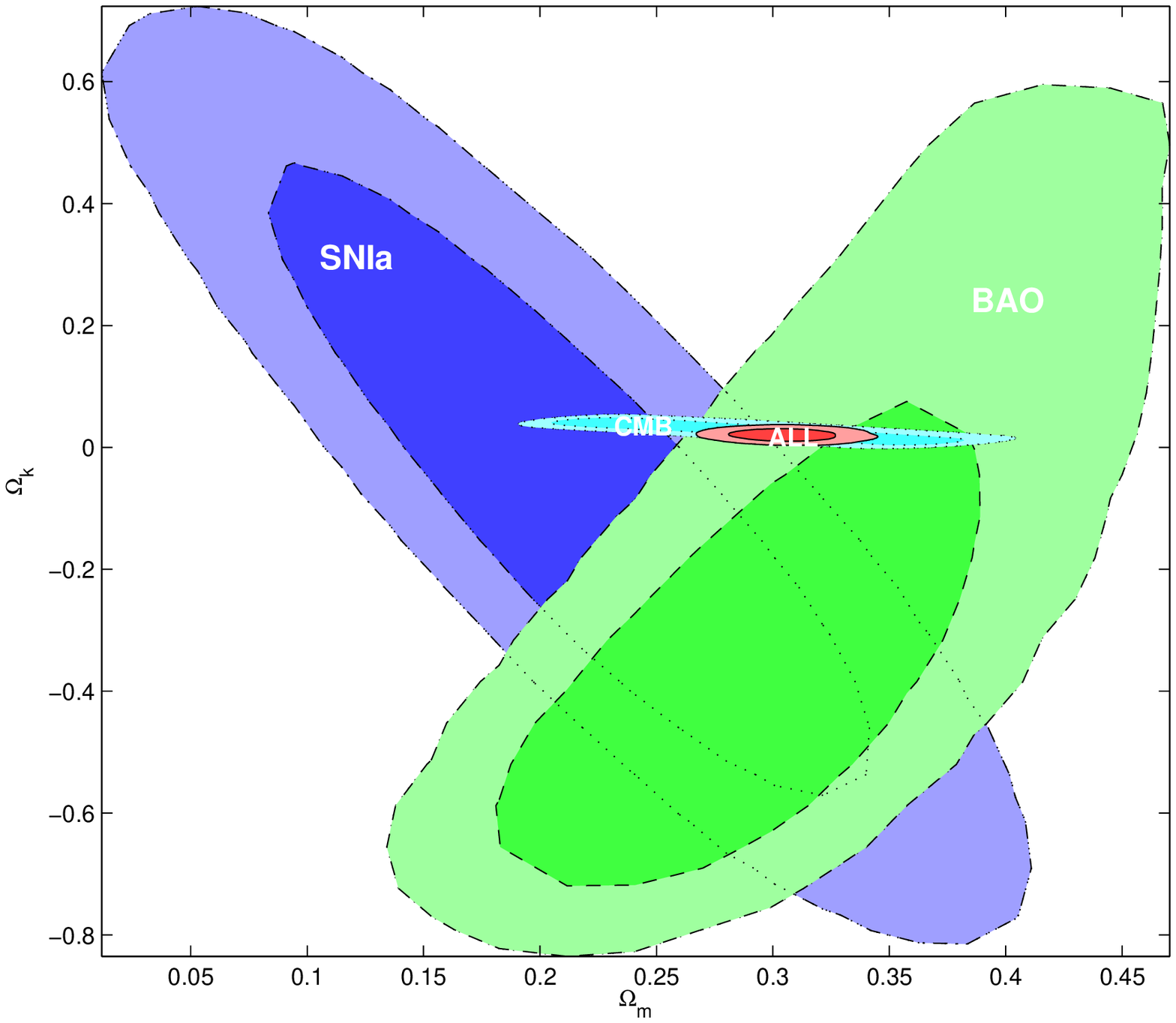}
\caption{The marginalized 1$\sigma$ and 2$\sigma$ contours of the DGP model parameters $\Omega_m$ and $\Omega_k$, derived from different data sets. ``ALL'' denotes the joint constraint including all the four data sets.}
\end{figure*}

\begin{figure*}
\includegraphics[width=7in]{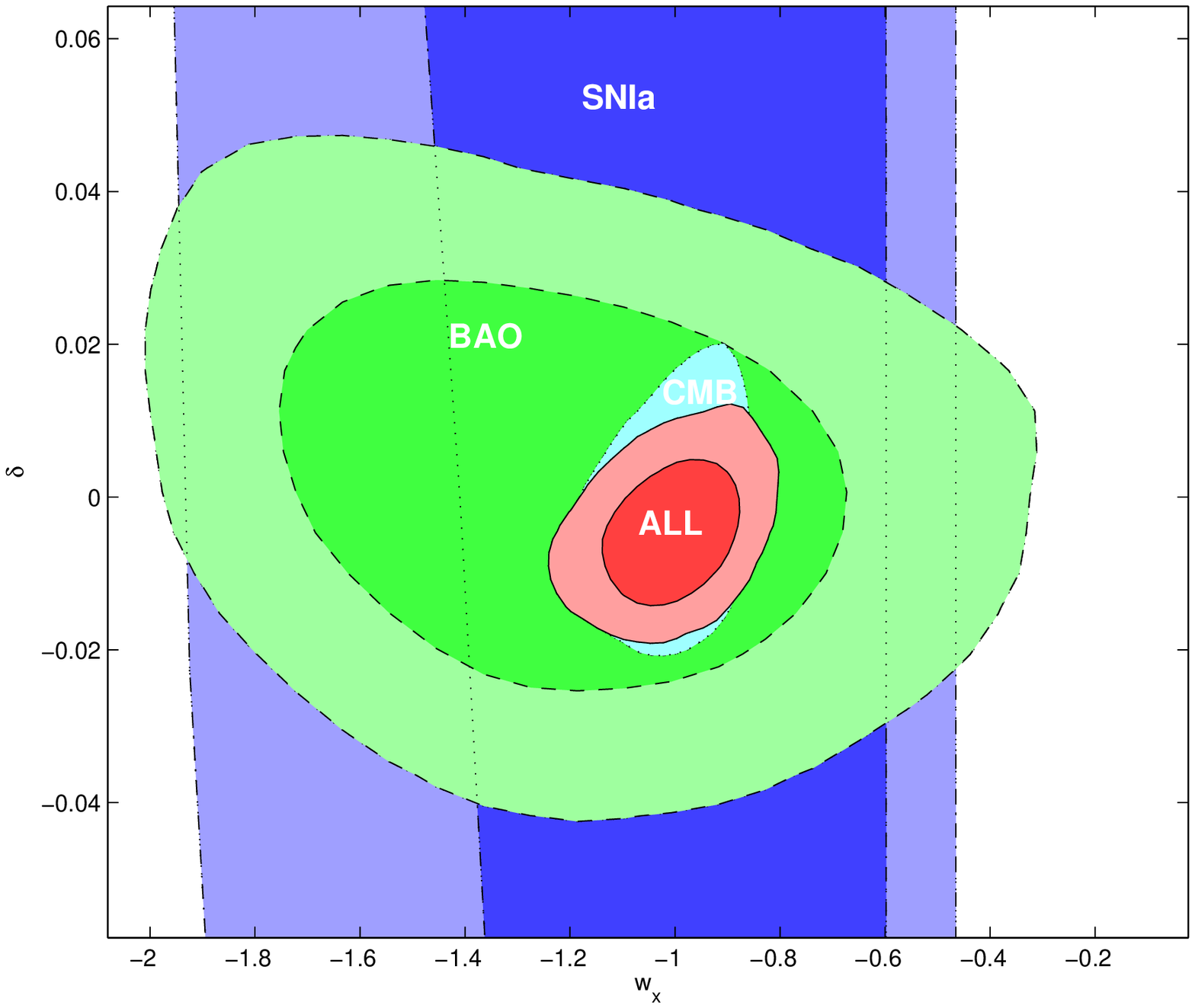}
\caption{The marginalized 1$\sigma$ and 2$\sigma$ contours of the IDE model parameters $\delta$ and $w_x$, derived from different data sets. ``ALL'' denotes the joint constraint including all the four data sets.}
\end{figure*}

\begin{figure*}
\includegraphics[width=7in]{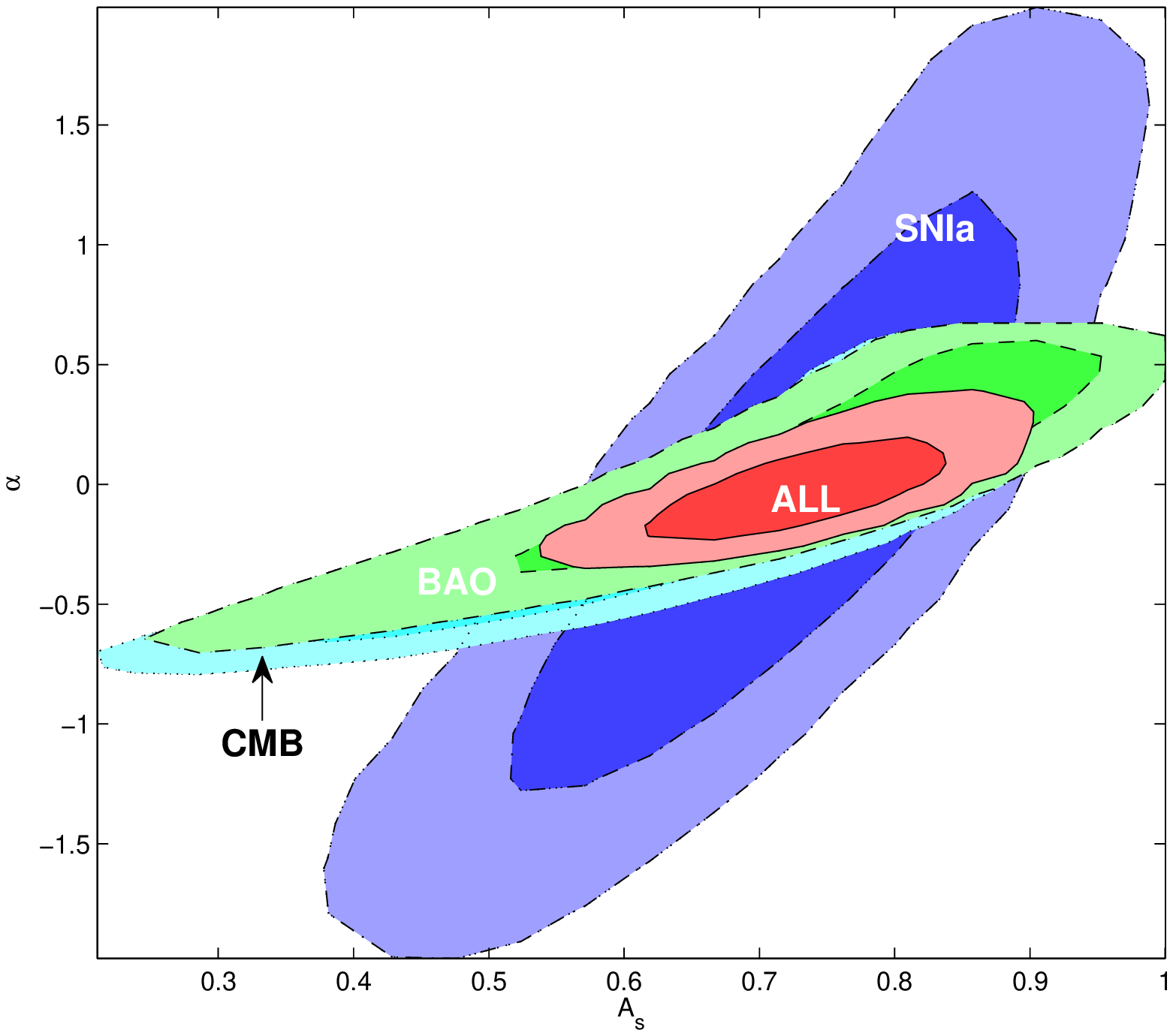}
\caption{The marginalized 1$\sigma$ and 2$\sigma$ contours of the GCG model parameters $A_s$ and $\alpha$, derived from different data sets. ``ALL'' denotes the joint constraint including all the four data sets.}
\end{figure*}

\begin{figure*}
\includegraphics[width=7in]{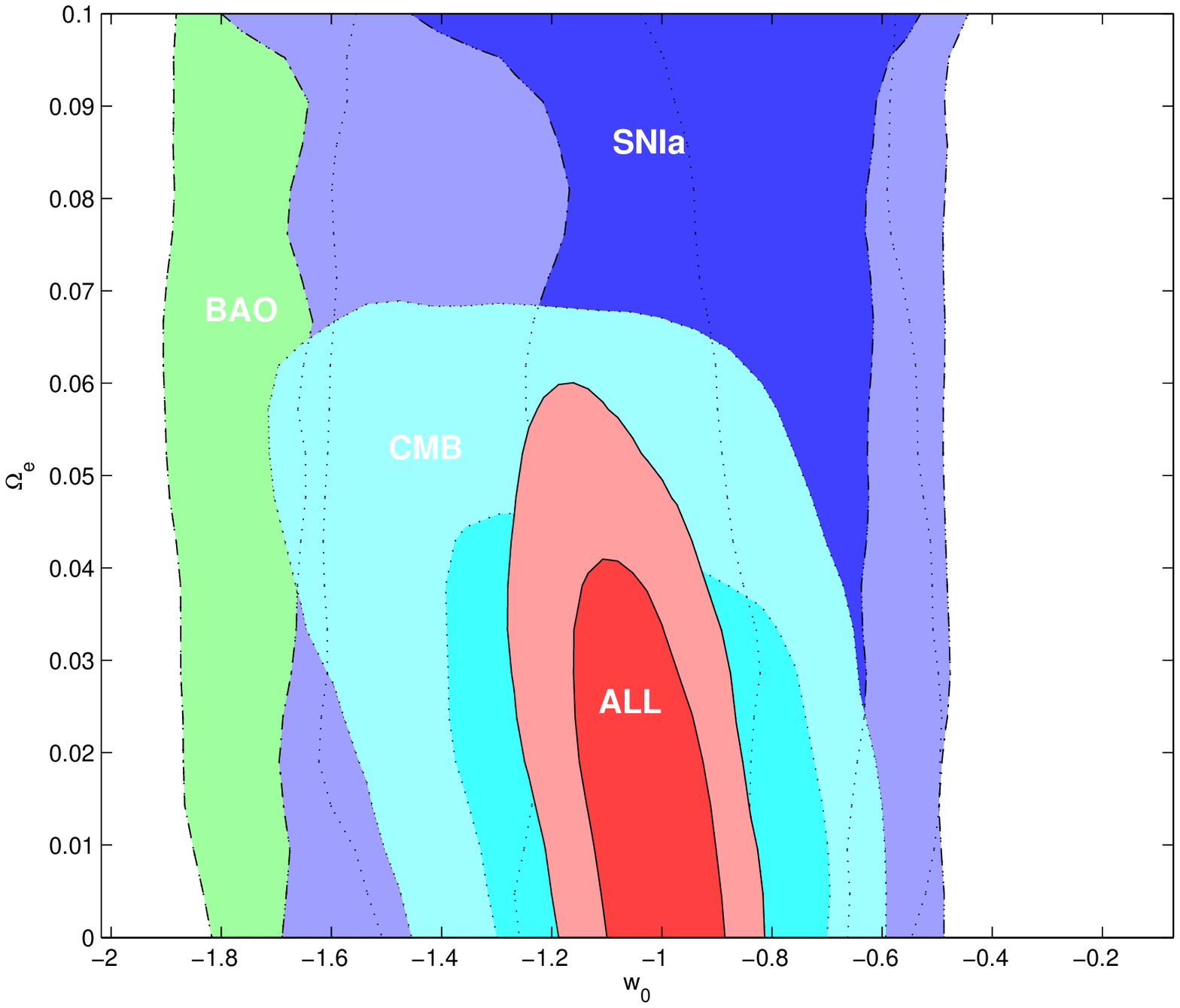}
\caption{The marginalized 1$\sigma$ and 2$\sigma$ contours of the EDE model parameters $w_0$ and $\Omega_e$, derived from different data sets. ``ALL'' denotes the joint constraint including all the four data sets.}
\end{figure*}

\begin{figure*}
$\begin{array}{cc}
\includegraphics[width=3.5in]{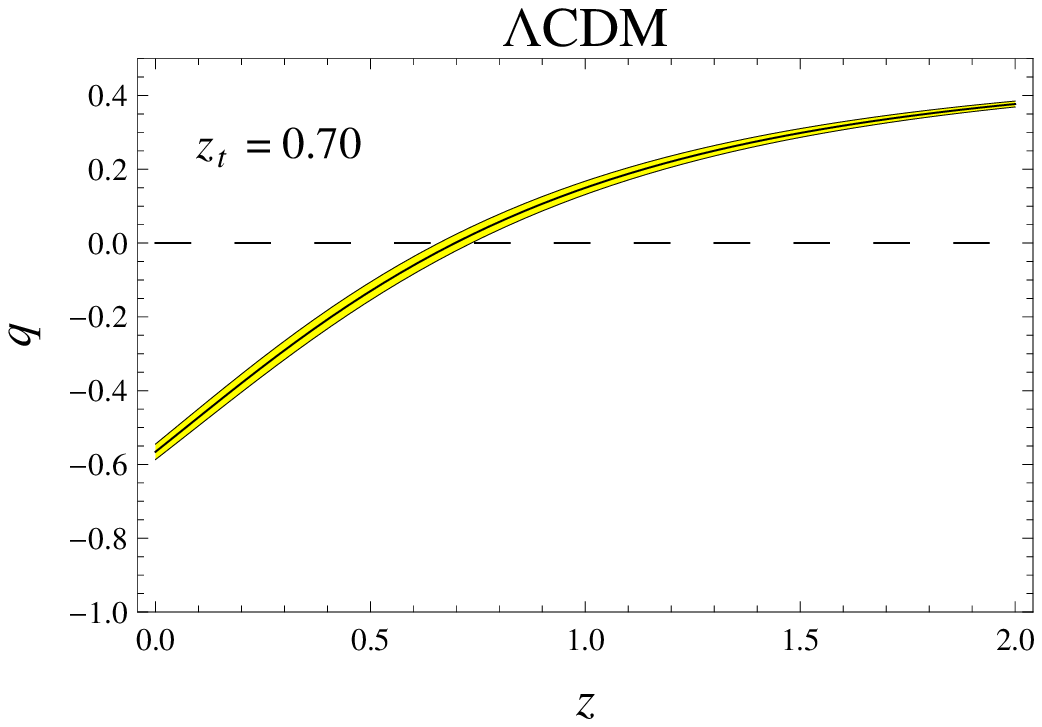}
\includegraphics[width=3.5in]{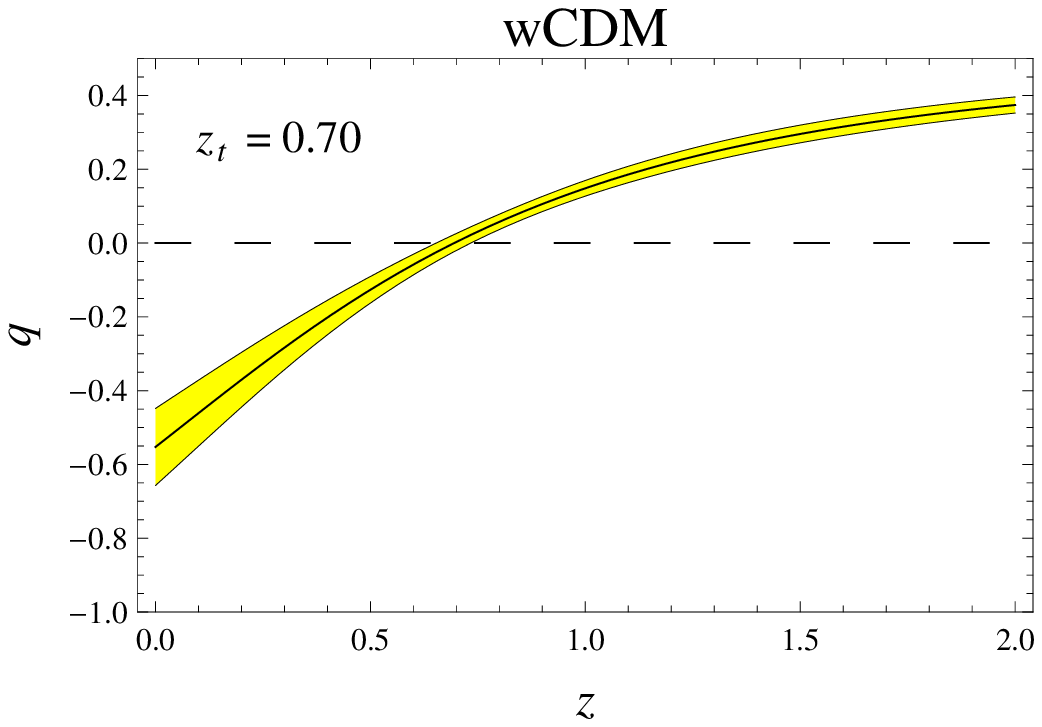}
\end{array}$
$\begin{array}{cc}
\includegraphics[width=3.5in]{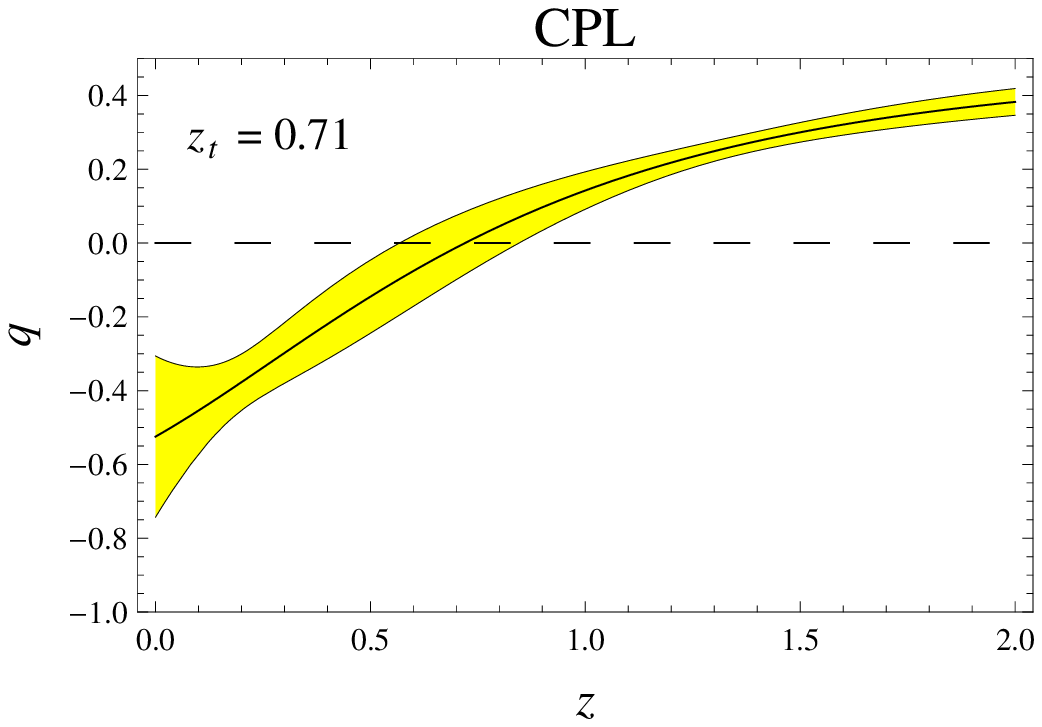}
\includegraphics[width=3.5in]{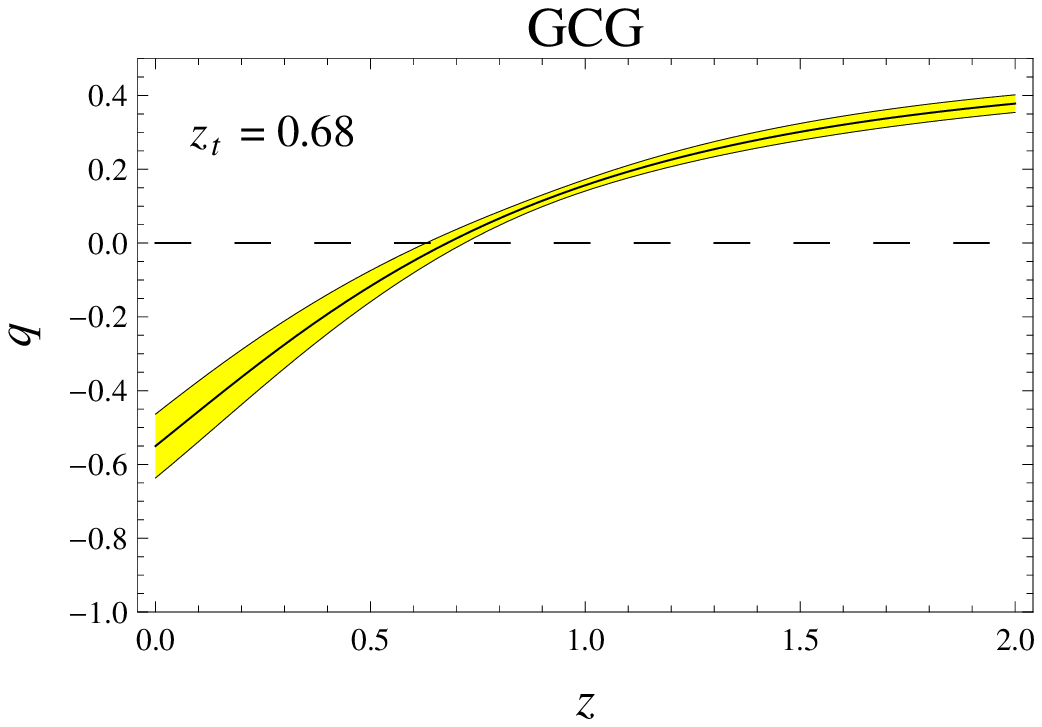}
\end{array}$
$\begin{array}{cc}
\includegraphics[width=3.5in]{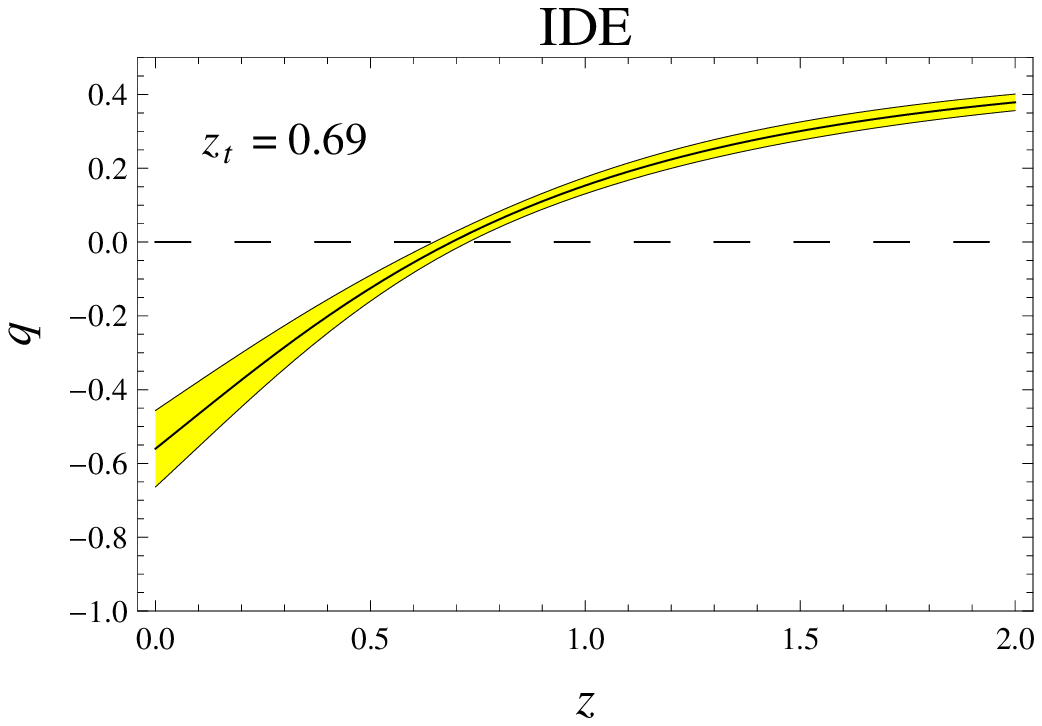}
\includegraphics[width=3.5in]{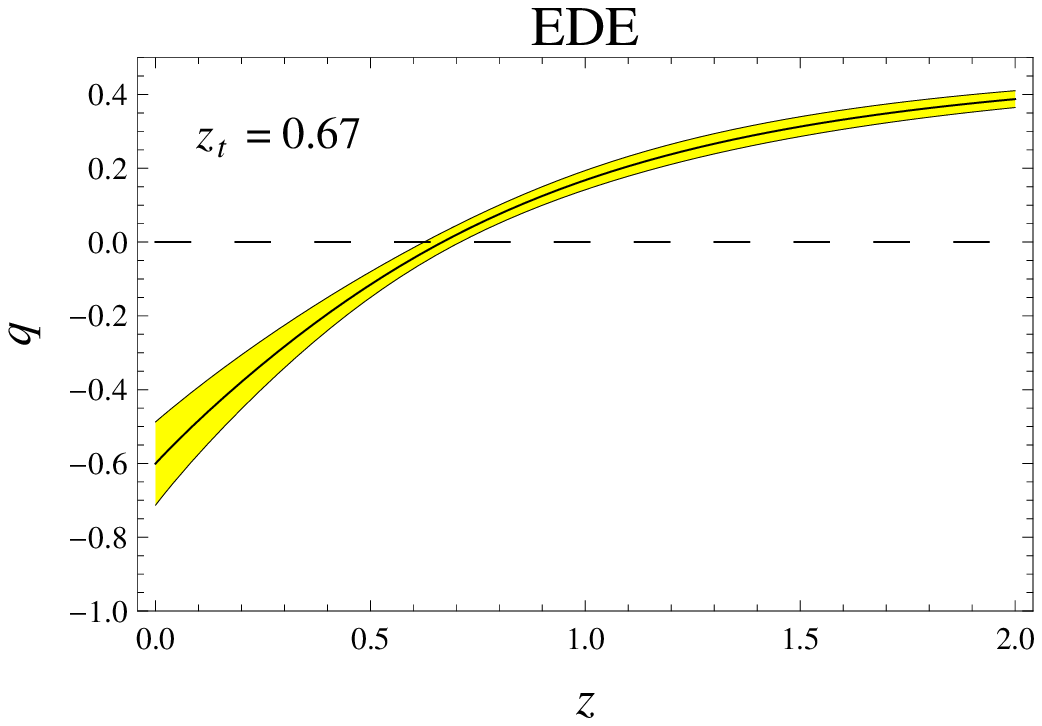}
\end{array}$
$\begin{array}{cc}
\includegraphics[width=3.5in]{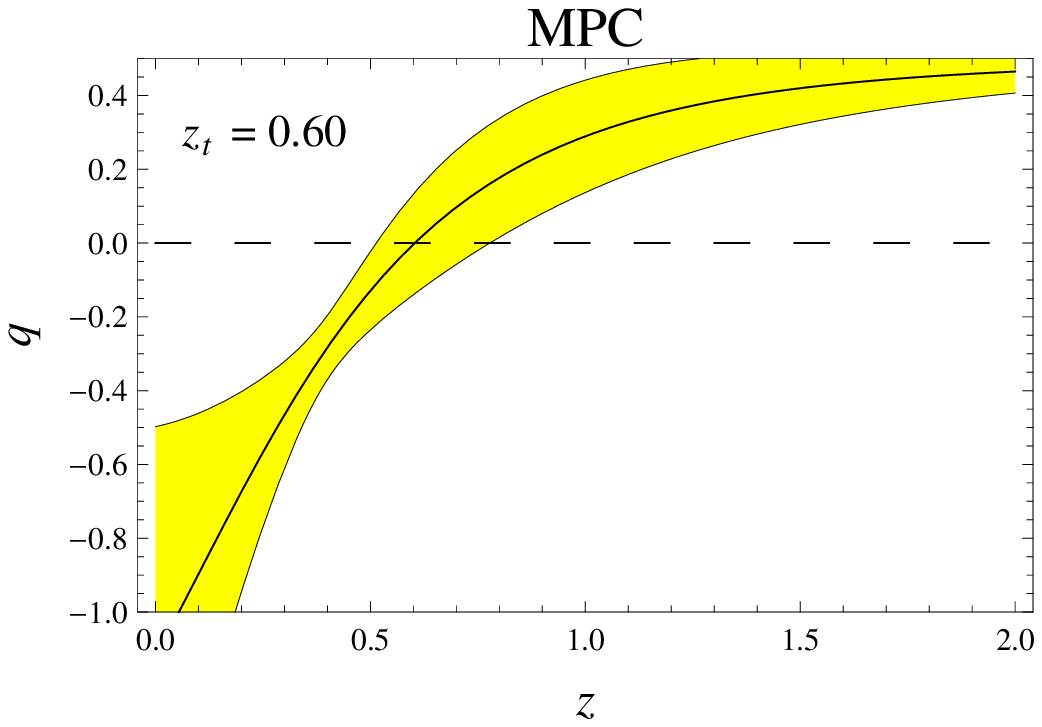}
\includegraphics[width=3.5in]{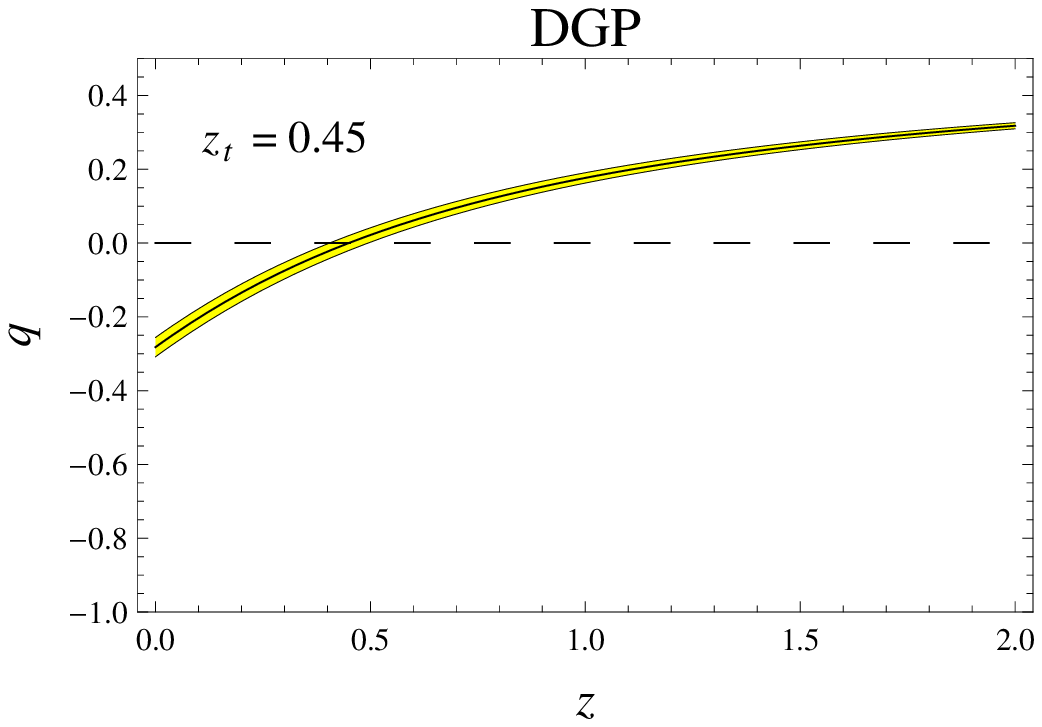}
\end{array}$
\caption{Evolution of the deceleration parameter $q$ for different cosmological models, the shaded regions show the 1 $\sigma$ uncertainties. The corresponding transition redshift $z_t$ is also given in each panel.}
\end{figure*}

\bibliographystyle{mn2e}
\bibliography{mybib}
\label{lastpage}
\end{document}

%% file: aas_macros.tex
\def\jnl@style{\it}
\def\ref@jnl#1{{\jnl@style#1}}

\def\aj{\ref@jnl{AJ}}                   
\def\actaa{\ref@jnl{Acta Astron.}}      
\def\araa{\ref@jnl{ARA\&A}}             
\def\apj{\ref@jnl{ApJ}}                 
\def\apjl{\ref@jnl{ApJ}}                
\def\apjs{\ref@jnl{ApJS}}               
\def\ao{\ref@jnl{Appl.~Opt.}}           
\def\apss{\ref@jnl{Ap\&SS}}             
\def\aap{\ref@jnl{A\&A}}                
\def\aapr{\ref@jnl{A\&A~Rev.}}          
\def\aaps{\ref@jnl{A\&AS}}              
\def\azh{\ref@jnl{AZh}}                 
\def\baas{\ref@jnl{BAAS}}               
\def\bac{\ref@jnl{Bull. astr. Inst. Czechosl.}}
\def\caa{\ref@jnl{Chinese Astron. Astrophys.}}
\def\cjaa{\ref@jnl{Chinese J. Astron. Astrophys.}}
\def\GReGr{\ref@jnl{Gen. Rel. Grav.}}   
\def\icarus{\ref@jnl{Icarus}}           
\def\IJMPD{\ref@jnl{Int. J. Modern Phys. D}}    
\def\jcap{\ref@jnl{J. Cosmology Astropart. Phys.}}
\def\jrasc{\ref@jnl{JRASC}}             
\def\memras{\ref@jnl{MmRAS}}            
\def\mnras{\ref@jnl{MNRAS}}             
\def\na{\ref@jnl{New A}}                
\def\nar{\ref@jnl{New A Rev.}}          
\def\pla{\ref@jnl{Phys. Lett. A}}       
\def\plb{\ref@jnl{Phys. Lett. B}}       
\def\pra{\ref@jnl{Phys.~Rev.~A}}        
\def\prb{\ref@jnl{Phys.~Rev.~B}}        
\def\prc{\ref@jnl{Phys.~Rev.~C}}        
\def\prd{\ref@jnl{Phys.~Rev.~D}}        
\def\pre{\ref@jnl{Phys.~Rev.~E}}        
\def\prl{\ref@jnl{Phys.~Rev.~Lett.}}    
\def\pasa{\ref@jnl{PASA}}               
\def\pasp{\ref@jnl{PASP}}               
\def\pasj{\ref@jnl{PASJ}}               
\def\rvmp{\ref@jnl{Rev. Mod. Phys.}}    
\def\rmxaa{\ref@jnl{Rev. Mexicana Astron. Astrofis.}}%
\def\qjras{\ref@jnl{QJRAS}}             
\def\skytel{\ref@jnl{S\&T}}             
\def\solphys{\ref@jnl{Sol.~Phys.}}      
\def\sovast{\ref@jnl{Soviet~Ast.}}      
\def\ssr{\ref@jnl{Space~Sci.~Rev.}}     
\def\zap{\ref@jnl{ZAp}}                 
\def\nat{\ref@jnl{Nature}}              
\def\iaucirc{\ref@jnl{IAU~Circ.}}       
\def\aplett{\ref@jnl{Astrophys.~Lett.}} 
\def\apspr{\ref@jnl{Astrophys.~Space~Phys.~Res.}}
\def\bain{\ref@jnl{Bull.~Astron.~Inst.~Netherlands}} 
\def\fcp{\ref@jnl{Fund.~Cosmic~Phys.}}  
\def\gca{\ref@jnl{Geochim.~Cosmochim.~Acta}}   
\def\grl{\ref@jnl{Geophys.~Res.~Lett.}} 
\def\jcp{\ref@jnl{J.~Chem.~Phys.}}      
\def\jgr{\ref@jnl{J.~Geophys.~Res.}}    
\def\jqsrt{\ref@jnl{J.~Quant.~Spec.~Radiat.~Transf.}}
\def\memsai{\ref@jnl{Mem.~Soc.~Astron.~Italiana}}
\def\nphysa{\ref@jnl{Nucl.~Phys.~A}}   
\def\physrep{\ref@jnl{Phys.~Rep.}}   
\def\physscr{\ref@jnl{Phys.~Scr}}   
\def\planss{\ref@jnl{Planet.~Space~Sci.}}   
\def\procspie{\ref@jnl{Proc.~SPIE}}   